\documentclass[10pt]{iopart}
\usepackage{iopams}  
\usepackage{graphicx} 
\usepackage[leftcaption]{./sidecap}

\def\vo{V} % volume
\def\su{A} % surface
\def\cu{C} % curve

\newcommand{\mb}[1]{\mbox{\boldmath $#1$}}
\renewcommand{\vec}[1]{\mbox{\boldmath $#1$}}
\newcommand{\half}{\leavevmode\kern.0em
            \raise.5ex\hbox{\tiny 1}\kern-.1em
                    /\kern-.15em\lower.25ex\hbox{\tiny 2}}

\begin{document}
%%%%%%%%%
	
\title{Towards relativistic simulations of  magneto-rotational core collapse}

\author{P Cerd\'a-Dur\'an$^{1,2}$ \& J A Font$^1$}
\address{$^1$ Departamento de Astronom\'ia y Astrof\'isica,
Universidad de Valencia, Dr. Moliner, 50, 
46100 Burjassot (Valencia), Spain}
\address{$^2$ Max-Plack-Institut f\"ur Astrophysik, 
Karl-Schwarzschild-Str. 1, 85741 Garching bei M\"unchen, Germany}
\ead{cerda@mpa-garching.mpg.de, j.antonio.font@uv.es}

\begin{abstract}
We present a new general relativistic hydrodynamics code specifically designed to study magneto-rotational, relativistic, stellar core collapse. The code is an extension of an existing (and thoroughly tested) hydrodynamics code, which has been applied in the recent past to study relativistic rotational core collapse. It is based on the conformally-flat  approximation of Einstein's field equations  and conservative formulations for the magneto-hydrodynamics equations. As a first step towards magneto-rotational core collapse simulations the code assumes a {\it passive} (test) magnetic field. The paper is focused on the description of the technical details of the numerical implementation, with emphasis on the magnetic field module. A number of code tests are presented and discussed, along with a representative core collapse simulation.
\end{abstract}

\submitto{\CQG}

%\maketitle

%%%%%%%%%%%
\section{Introduction}
%%%%%%%%%%%

During the last decades many efforts have been taken to numerically simulate the collapse of stellar cores (see e.g.~\cite{LRfryer,dimmelmeier_05_a,woosley05} and references therein). A myriad of difficulties arise for its modelling as many physical effects are involved in a process whose dynamics is highly nonlinear. Analytic approaches are of limited use  and numerical simulations are needed to solve the coupled system of equations describing the fluid motion, the spacetime evolution, the dynamics of the magnetic fields, and the neutrino transport. Such simulations are driving progress in the field despite the limited knowledge on relevant {\it physical} issues such as realistic precollapse stellar models (including rotation and magnetic field strength and distribution) or microphysical, finite-temperature equation of state (EOS), as well as {\it numerical} limitations involved in the challenging task of accounting for Boltzmann neutrino transport, multidimensional hydrodynamics, and relativistic gravity.

Only very recently the first multidimensional simulations of relativistic rotational core collapse have become possible, thanks to the use of conservative formulations of the hydrodynamics equations and advanced numerical methodology, as well as long-term stable formulations of  Einstein's equations (or accurate enough approximations)~\cite{dimmelmeier_02_b,ss04} (see also~\cite{baiotti}). On the other hand, the incorporation of magnetic fields and the MHD equations in numerical codes to further  improve the realism of  such  simulations in general relativity, is currently an emerging field where significant progress is bound to be soon achieved (see~\cite{duez06,shibata06}). 

Neutron stars have intense magnetic fields ($\sim 10^{12}-10^{13}$ G) or even larger at birth. The presence of such magnetic fields renders magneto-rotational core collapse simulations mandatory. In recent years, an increasing number of authors have performed axisymmetric  simulations (within the so-called ideal MHD limit) employing Newtonian treatments of the magneto-hydrodynamics, the gravity, and of the microphysics (see~\cite{martin} and references therein). The weakest point of all existing  simulations to date is the fact that the strength and distribution of the initial magnetic field in the core are basically unknown. The available simulations show that for weak initial fields ($\leq 10^{11}$ G, which is the most relevant case, astrophysics-wise) there are no major differences in the collapse dynamics nor in the resulting gravitational wave signal, when comparing with purely hydrodynamical simulations. However,  strong initial fields ($\geq 10^{11}$ G) manage to slow down the core efficiently (leading even to retrograde rotation in the proto-neutron star (PNS)) which causes qualitatively different dynamics and gravitational wave signals. 

This paper presents a general relativistic magneto-hydrodynamics (GRMHD) code designed to study magneto-rotational, relativistic, stellar core collapse. Our GRMHD code is an extension of the hydrodynamics code developed by~\cite{dimmelmeier_02_a,dimmelmeier_02_b}, where magnetic fields are included following the approach laid out in~\cite{anton06}. Einstein's equations are formulated using the conformally flat condition (CFC), which has proved very accurate for studying rotational core collapse~\cite{dimmelmeier06}.  Leaving aside  radiation transport, whose effects on core collapse dynamics and gravitational radiation are dramatic and its numerical modelling is a challenge in itself (see~\cite{woosley05} and references therein), we adopt here as a first step towards simulating realistic magneto-rotational core collapse the {\it passive} magnetic field approximation, which is  justifiable for the magnetic field values present in this scenario. The paper focuses on describing the details of the numerical schemes  we use and the tests performed with the code. A representative magneto-rotational core collapse simulation is also presented and briefly discussed. A parameter-space survey of the magneto-rotational core collapse scenario will be presented elsewhere. We note that during the development of this work we have been made aware of a similar study reported in~\cite{shibata06b}.

%%%%%%%%%%%%%%%%%%%%%%%%%%%%%%%%%%%%%%%%%%%%%%%%%
\section{Mathematical and physical framework}
%%%%%%%%%%%%%%%%%%%%%%%%%%%%%%%%%%%%%%%%%%%%%%%%%

We adopt the $3+1$ formalism to foliate the spacetime into spacelike hypersurfaces. In this formulation the metric reads $ds^2 = g_{\mu\nu}dx^{\mu}dx^{\nu}=- \alpha^2 dt^2 + \gamma_{ij} (dx^i + \beta^i dt)(dx^j + \beta^j dt)$, where $\alpha$ is the lapse function, $\beta^i$ the shift vector, and $\gamma_{ij}$ the spatial three-metric induced in each hypersurface.  From the energy-momentum tensor $T^{\mu\nu}$ it is possible to build the following quantities: $E \equiv n^{\mu} n^{\nu} T_{\mu\nu} = \alpha^2 T^{00}$, $S_i \equiv - \perp^{\mu}_{i} n^{\nu} T_{\mu\nu} = -\frac{1}{\alpha} (T_{0i} - T_{ij} \beta^j)$, and $S_{ij}  \equiv \perp^{\mu}_i \perp^{\nu}_j T_{\mu\nu} = T_{ij}$, using the projection operator $\perp^{\mu}_{\nu}$ and the unit four-vector  $n^{\mu}$ normal to the hypersurface.

%%%%%%%%%%%%%%%%%%%%%%%%%%%%%%%%%%%%%%%%%%%%%%%%
\subsection{The CFC approximation}
%%%%%%%%%%%%%%%%%%%%%%%%%%%%%%%%%%%%%%%%%%%%%%%%

To solve Einstein's field equations we use the CFC approximation~\cite{wilson_96_a}. In this approach, the three-metric in the ADM gauge $\gamma_{ij} = \phi^4 \hat{\gamma}_{ij} + h^\mathrm{TT}_{ij}$, is assumed to be conformally flat, $\gamma_{ij} = \phi^4 \hat\gamma_{ij}$. In these expressions  $ \phi $ is the conformal factor, $ h^\mathrm{TT}_{ij} $ is transverse and traceless, and $\hat{\gamma}_{ij}$ is the flat three-metric. Note that the ADM gauge choice implies the maximal slicing condition in which the trace of the extrinsic curvature vanishes. Under the CFC approximation, Einstein's field equations can be written as a system of five elliptic equations
\begin{eqnarray}
\hat{\Delta} \phi &= - 2 \pi \phi^5 \left ( E + \frac{K_{ij}K^{ij}}{16 \pi} \right ),  \label{eq:cfc1}\\
\hat{\Delta} (\alpha\phi) &= 2\pi \alpha \phi^5 \left (
E + 2 S + \frac{7 K_{ij}K^{ij}}{16 \pi} 
\right ), \label{eq:cfc2}\\
\hat{\Delta} \beta^i &= 16 \pi \alpha \phi^4 S^i 
+ 2 \phi^{10} K^{ij} \hat{\nabla}_j \left( \frac{\alpha}{\phi^6} \right)
- \frac{1}{3} \hat{\nabla}^i \hat{\nabla}_k \beta^k, \label{eq:cfc3}
\end{eqnarray}
where $\hat{\Delta}$ and $\hat{\nabla}$ are the Laplacian and divergence operators in flat spacetime, $S\equiv \gamma^{ij}S_{ij}$, and  $K^{ij}$ is the extrinsic curvature.

%%%%%%%%%%%%%%%%%%%%%%%%%%%%
\subsection{General relativistic magneto-hydrodynamics}
%%%%%%%%%%%%%%%%%%%%%%%%%%%%

For a perfect fluid endowed with an electromagnetic field the energy-momentum tensor is the sum of a fluid part and of a electromagnetic field part
\begin{eqnarray}
T^{\mu\nu}_{\rm Fluid} &= \rho h u^{\mu} u^{\nu} + P g^{\mu\nu}, \qquad
T^{\mu\nu}_{\rm EM} = F^{\mu\lambda}F^{\nu}_{\lambda} 
- \frac{1}{4} g^{\mu\nu} F^{\lambda\delta}F_{\lambda\delta},
\end{eqnarray}
where $\rho$ is the rest-mass density, $h= 1 + \epsilon + P/\rho$ is the relativistic enthalpy, $\epsilon$ is the specific internal energy, $P$ is the pressure, and $u^{\mu}$ is  the four-velocity of the fluid. The electromagnetic  tensor $F^{\mu\nu} = U^{\mu} E^{\nu} - U^{\nu} E^{\mu} - \varepsilon^{\mu\nu\lambda\delta}U_{\lambda} B_{\delta}$, and its dual $^{*}F^{\mu\nu}$ can be expressed in terms of the electric field $E^{\mu} = F^{\mu\nu}U_{\nu}$ and the magnetic field $B^{\mu} = \,^{*}F^{\mu\nu} U_{\nu}$ measured by an observer  with four-velocity $U^{\mu}$. We denote by $e^{\mu}$ and $b^{\nu}$ the electric and magnetic fields, respectively, measured by a comoving observer $u^{\mu}$.

%%%%%%%%%%%%%%%%%%
\subsubsection{Maxwell's equations.}
%%%%%%%%%%%%%%%%%%

The equations governing the evolution of the electro-magnetic fields are Maxwell's equations, which can be written in terms of the Faraday tensor as $^{*}F^{\mu\nu}_{\quad;\nu} = 0$ and $F^{\mu\nu}_{\quad;\nu} = 4\pi \mathcal{J}^{\mu}$, where $\mathcal{J}^{\mu}$ is the electric four-current. Under the assumption that Ohm's law is fulfilled, the latter reads $\mathcal{J}^{\mu} = \rho_{\mathrm{q}} u^{\mu} + \sigma e^{\mu}$, where $\rho_{\mathrm{q}}$ is the proper charge density  and $\sigma$ is the electric conductivity. Maxwell's equations can be simplified if the fluid is a perfect conductor ($\sigma \to \infty$). In this case, to keep the current finite, $e^{\mu}$ must vanish. This case corresponds to the so-called ideal MHD condition. Under this condition the four-vector electric field $E^{\mu}$ can be expressed in terms of the four-vector magnetic field $B^{\mu}$, and, thus, only  equations for $B^i$ are needed. It is  convenient to choose as observer the Eulerian observer, $U^{\mu}=n^{\mu}$, for which the temporal component of the electric field vanishes, $E^{\mu} = (0, -\varepsilon_{ijk}v^j B^k)$. In this case the first set of Maxwell's equations reduce to the divergence-free condition plus the induction equation for the magnetic field
\begin{eqnarray}
\hat{\nabla}_i B^{*i}&= 0 ,\qquad\qquad
\frac{\partial B^{*i}}{\partial t} 
&= \hat{\nabla}_j
(v^{*i} B^{*j} - v^{*j} B^{*i}),
\label{eq:induction}
\end{eqnarray}
where $B^{*i} \equiv \sqrt{\bar{\gamma}} B^i$ and $v^{*i} \equiv \alpha \hat{v}^i \equiv \alpha v^i - \beta^i$,  $v^i$ being the three-velocity as measured by the Eulerian observer. Here, $\bar{\gamma}$ denotes the ratio of the determinants of the three-metric and the flat three-metric, $\bar{\gamma} = \gamma / \hat{\gamma}$. 

%%%%%%%%%%%%%%%%%%%%%%
\subsubsection{Magnetic flux conservation.}
%%%%%%%%%%%%%%%%%%%%%%

The total magnetic flux through a closed surface $\hat{\su}$ enclosing a volume $\hat{\vo}$ can be calculated as a surface integral of the ``starred'' magnetic field as
\begin{eqnarray}
\Phi_{\rm T} = \oint_{\hat{\su}=\partial \hat{\vo}} \vec{B}^{*} \cdot d\vec{\hat{\su}}
= \int_{\hat{\vo}} \vec{\hat{\nabla}} \cdot \vec{B}^* d\hat{\vo} = 0,
\end{eqnarray}
after applying Gauss theorem and the magnetic field divergence-free constraint. Quantities in boldface correspond to three-vectors. The scalar product ($\cdot$), and the cross product ($\times$) used below, are defined with respect to the flat three-metric $\hat{\gamma}_{ij}$. This equation implies that no source of magnetic flux exists inside the volume $\hat{\vo}$ and, therefore, the magnetic flux is a conserved quantity as  $\frac{\partial \Phi_{\rm T}}{\partial t} = 0$.

In addition, if we consider a generic surface $\hat{\su}$ (without the restriction of having to enclose a volume), the time variation of the magnetic flux through the surface is
\begin{eqnarray}
\hspace{-0.8cm}\frac{\partial \Phi}{\partial t} 
&= \frac{\partial}{\partial t}\int_{\hat{\su}} \vec{B}^{*} \cdot d \vec{\hat{\su}}
= \int_{\hat{\su}} 
\left[ \vec{\hat{\nabla}} \times \left( \vec{v}^* \times \vec{B}^* \right) \right] \cdot d\vec{\hat{\su}} 
= -\oint_{\hat{\cu}=\partial \hat{\su}} \vec{E}^* \cdot d\vec{\hat{l}},
\label{eq:mflux_suface}
\end{eqnarray}
where we have used the induction equation (\ref{eq:induction}) and Stokes theorem to transform the surface integral into a line integral along the curve $\hat{\cu}$ enclosing $\hat{\su}$, and the equality 
$\vec{E}^* = \vec{v}^* \times \vec{B}^*$.
The two properties just inferred allow us to design a numerical algorithm to solve the induction equation and the divergence constraint in a way that ensures the conservation of the magnetic flux.

%%%%%%%%%%%%%%%%%%
\subsubsection{Conservation laws.}
%%%%%%%%%%%%%%%%%%

In the ideal MHD limit the energy-momentum tensor of the electromagnetic field can be written in terms of the magnetic field $b^{\mu}$ measured by a comoving observer. The total energy-momentum tensor is thus given by
\begin{equation}
T^{\mu\nu} =  T^{\mu\nu}_{\rm Fluid} + T^{\mu\nu}_{\rm EM}
= \left(\rho h + b^2 \right)\, u^{\mu} u^{\nu} 
+ \left(P + \frac{b^2}{2}\right)  g^{\mu\nu} - b^{\mu} b^{\nu},
\label{eq:tmunu_grmhd}
\end{equation}
where $b^2 = b_{\mu}b^{\mu}$. We define the magnetic pressure $P_{\rm mag}=b^2/2$ and the specific magnetic energy $\epsilon_{\rm mag} = b^2 /2 \rho $, as their effects on the dynamics are similar to those played by the pressure and the specific internal energy. The evolution of the magneto-fluid is determined by the conservation law of the energy-momentum $T^{\mu\nu}_{\quad;\mu}  = 0$ and the continuity equation $J^{\mu}_{\,\,\,\,;\mu}  = 0$, for the rest-mass current $J^{\mu} = \rho u^{\mu}$. Following the procedure laid out in~\cite{anton06}, the magnetic field can be accounted for by choosing the conserved quantities in a similar way to the purely hydrodynamical case~\cite{banyuls}
\begin{eqnarray}
     D &=& \rho W, \\
  S_i  &=&  (\rho h + b^2)W^2 v_i - \alpha b_i b^0, \\
  \tau &=&  E - D = (\rho h +b^2) W^2 - \left(P + \frac{b^2}{2}\right) - \alpha^2 (b^0)^2 - D.
\end{eqnarray}
With this choice, the system of conservation equations for the fluid and the induction equation for the magnetic field can be cast as a first-order, flux-conservative, hyperbolic system, as 
\begin{equation}
  \frac{1}{\sqrt{- g}} \left[
  \frac{\partial \sqrt{\gamma} \mb{U}}{\partial t} +
  \frac{\partial \sqrt{- g} \mb{F}^i}{\partial x^i} \right] = \mb{Q},
  \label{eq:hydro_conservation_equation}
\end{equation}
with the state vector, flux vector, and source vector given, respectively, by
\begin{eqnarray}
  \mb{U} & =& [D, S_j, \tau, B^k], \\
  \mb{F}^i & =& \left[
  D \hat{v}^i
  ,S_j \hat{v}^i + \delta^i_j \left(P + \frac{b^2}{2}\right) -  \frac{b_j B^i}{W} 
  \right.\nonumber \\&&\left.
  , \tau \hat{v}^i + \left(P + \frac{b^2}{2}\right) v^i - \alpha \frac{b^0 B^i}{W}
  , \hat{v}^i B^k - \hat{v}^k B^i
  \right], \label{eq:mhydro_conservation_equation_vectors_flux} \\
  \mb{Q} & =& \left[ 0, \frac{1}{2} T^{\mu \nu}
  \frac{\partial g_{\mu \nu}}{\partial x^j},
  \alpha \! \left( \! T^{\mu 0} \frac{\partial \ln \alpha}{\partial x^\mu} -
  T^{\mu \nu} {\it \Gamma}^0_{\mu \nu} \! \right) , 0^k \! \right] \!.
  \label{eq:mhydro_conservation_equation_vectors}
\end{eqnarray}%
We note that these expressions contain components of the magnetic field measured by both, a comoving observer and an Eulerian observer. The two are related by
\begin{eqnarray}
b^0 &= \frac{W B^i v_i}{\alpha}, \qquad\qquad
b^i &= \frac{B^i + \alpha b^0 u^i}{W}.
\end{eqnarray}

%%%%%%%%%%%%%%%%%%
\subsubsection{Hyperbolic structure.}
%%%%%%%%%%%%%%%%%%

Anticipating the numerical methods that we use to solve the conservation equations (\ref{eq:hydro_conservation_equation}), we need to know the wave structure of the hyperbolic system of equations. The associated flux-vector Jacobians in every direction are $7 \times 7$ matrices, and the solution of the eigenvalue problem \cite{antonphd,anton06} leads to seven types of waves, which may appear when solving the Riemann problem for each direction~$i$: the entropic wave $\lambda^{i}_e = \alpha v^{i} - \beta^{i}$, the Alfven waves $\lambda^i_{a\pm}= (b^i\pm\sqrt{\rho h + B^2} u^i)/(b^0 \pm \sqrt{\rho h + B^2}u^0)$, and the  magnetosonic waves. There is no analytic expression for the latter, which must be computed as the solution of a quartic equation (see~\cite{anton06} for details). Among the magnetosonic waves, the two solutions with maximum and minimum speeds are called {\it fast} magnetosonic waves $\lambda^i_{f\pm}$, and the two solutions in between are the so-called {\it slow} magnetosonic waves $\lambda^i_{s\pm}$. The seven waves can be ordered as follows
%
%\begin{equation}
$\lambda^i_{f-} \le \lambda^i_{a-} \le \lambda^i_{s-} \le \lambda^i_{e} \le \lambda^i_{s+} 
\le \lambda^i_{a+} \le \lambda^i_{f+}$.
%\end{equation}

%%%%%%%%%%%%%%%%%%%%%%%%
\subsubsection{The passive field approximation.}
%%%%%%%%%%%%%%%%%%%%%%%%

In the collapse of {\it weakly magnetized} stellar cores, it is a good approximation to consider that the magnetic field entering in the energy-momentum tensor of (\ref{eq:tmunu_grmhd}) is negligible when compared with the fluid part, i.e. $P_{\rm mag} \ll P$, $\epsilon_{\rm mag} \ll \epsilon$, and that the components of the anisotropic term of $T^{\mu \nu}$ satisfy $b^{\mu}b^{\nu} \ll \rho h u^{\mu} u^{\nu} + P g^{\mu \nu}$. With such simplifications the remaining system of equations comprises the hydrodynamics equations (with no magnetic field) and the induction equation. In this case, the magnetic field evolution does not affect the dynamics of the fluid, but it does depend on the evolution of the matter due to the presence of the velocity components in the induction equation.

Such ``test magnetic field'' (or passive field) approximation is employed in the core collapse simulations reported in this paper. In this  approximation the seven eigenvalues of the GRMHD Riemann problem reduce to three
\begin{eqnarray}
&\lambda^{i}_{0{\, \rm hydro}} = \lambda^i_e = \lambda^i_{a\pm} = \lambda^i_{s\pm} ,  \qquad\qquad
\lambda^i_{\pm{\, \rm hydro}} =\lambda^i_{f\pm} , 
\end{eqnarray}
where $\lambda^{i}_{0{\, \rm hydro}}$ and $\lambda^i_{\pm{\, \rm hydro}}$ are the eigenvalues of the Jacobian matrices of the hydrodynamics equations (see \cite{font_03_a}).

%%%%%%%%%%%%%%%
\section{Numerical framework}
%%%%%%%%%%%%%%%

The GRMHD numerical code presented in this paper is based on the hydrodynamics code described in \cite{dimmelmeier_02_a,dimmelmeier_02_b}, and on its extension discussed in \cite{cerda05}. The code performs the coupled evolution of the equations governing the dynamics of the spacetime, fluid, and magnetic fields in general relativity. The equations are implemented in the code using spherical polar coordinates $ \{ t, r, \theta, \varphi \} $, assuming axisymmetry with respect to the rotation axis and equatorial plane symmetry at $ \theta = \pi / 2 $. For the various types of PDEs implemented in the code we use the most appropriate numerical methods, which will be described next. 

%%%%%%%%%%%%%%%%%%%%
\subsection{The hydrodynamics solver}
\label{sec:hydro_solver}
%%%%%%%%%%%%%%%%%%%%

Since we adopt the passive magnetic field approximation, the fluid evolution is not affected by the magnetic field. Such evolution is performed using a high-resolution shock-capturing (HRSC) scheme which numerically integrates a subset of equations in system~(\ref{eq:hydro_conservation_equation}), the one corresponding to the purely hydrodynamical variables ($D, S_i, \tau$). HRSC methods ensure numerical conservation of physically  conserved quantities and a correct treatment of discontinuities such as shocks (see e.g. \cite{font_03_a} and references therein). Following \cite{anton06} we use a linear reconstruction procedure with a {\it minmod} slope limiter, which yields second order accuracy in space. Correspondingly, the time update of the state vector $ \mb U $ is done using the method of lines in combination with a second-order accurate Runge--Kutta scheme. The numerical fluxes at cell interfaces are obtained  using  ``incomplete'' approximate Riemann solvers, i.e. solvers that do not need the full characteristic information of the system. This kind of solvers are particularly useful in GRMHD, where the full set of eigenspeeds of the flux-vector Jacobians is not known in a closed form \cite{anton06}. We have implemented the HLL single-state solver of \cite{harten83} and the symmetric scheme of \cite{kt00} (KT). Both solvers yield results with an accuracy comparable to complete Riemann solvers (with the full characteristic information), as shown in simulations involving purely hydrodynamical special relativistic flows \cite{lucas04} and general relativistic flows in dynamical spacetimes \cite{shibata_font05}. Test of both solvers in GRMHD have been reported recently by \cite{anton06}. 
 
%%%%%%%%%%%%%%%%%%
\subsection{Magnetic field evolution}
%%%%%%%%%%%%%%%%%%

The magnetic field evolution needs to be performed in a different way to the rest of the conservation equations, since the physical meaning of the corresponding conservation equation is different. Although the induction equation can be written in a flux conservative way, a supplementary condition for the magnetic field has to be given (the divergence constraint), and it has to be fulfilled at each time iteration. As shown before, the physical meaning of these two equations is the conservation of the magnetic flux in a closed volume, in our case each numerical cell. Therefore, an appropriate numerical scheme has to be used which takes full profit of such conservation law. Among the numerical schemes that satisfy this property (see \cite{toth00} for a review), the constrained transport (CT) scheme \cite{evans88} has proved to be adequate to perform accurate simulations of magnetized flows. Our particular implementation of this scheme has been adapted to the spherical polar coordinates used in the code.

%%%%%%%%%%%%%%%%%%%%%%%%%%%%%
\subsubsection{CT scheme in spherical polar coordinates.}
%%%%%%%%%%%%%%%%%%%%%%%%%%%%%

\begin{SCfigure*}[4][t]
\includegraphics[width=.42\textwidth]{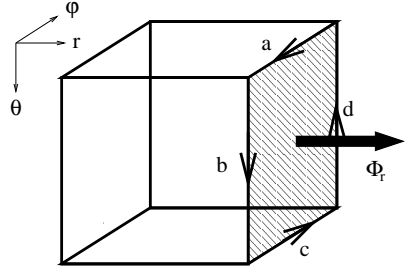}
  \caption{Schematic representation of the numerical cell. The time derivative of the magnetic fluxes $\Phi_i$ over the $r$ interface, the $\theta$ interface, and the $\varphi$ interface can be written as line integrals along the corresponding closed path a-b-c-d. Only the first case is shown in the plot.}
  \label{fig:fluxes}
\end{SCfigure*}

In order to implement the CT scheme in our numerical code we have to analyze how the magnetic flux behaves at the surface of a numerical cell. Therefore, we apply (\ref{eq:mflux_suface}) for the magnetic flux evolution in the interfaces along the direction of each coordinate (see figure~\ref{fig:fluxes}). To do this we assume that $B^{*i}$ is constant over each cell surface, and $E^*_i$ is constant along each cell edge. In the $r$ direction this yields
\begin{eqnarray}
\frac{\partial \Phi_r}{\partial t} 
&=& \Delta \hat{\su}_r \ \frac{\partial}{\partial t} B^{* r} 
= [E^*_{\varphi} \Delta \hat{l}_{\varphi} ]_a 
- [E^*_{\varphi} \Delta \hat{l}_{\varphi} ]_c ,
\end{eqnarray}
keeping in mind that the axisymmetry condition imposes that $[E^*_{\theta} \Delta \hat{l}_{\theta} ]_b = [E^*_{\theta} \Delta \hat{l}_{\theta} ]_d$. We define $\Delta \hat{\su}_i \equiv \int d \hat{\su}_i $ as the surface of the cell interface perpendicular to the $i$ direction and $\Delta \hat{l}_i \equiv \int d\hat{l}_i$ as the length of the cell edge in the $i$ direction. In the $\theta$ and $\varphi$ directions similar relationships can be found.

We represent all these quantities in the numerical grid, which has a total number of $n_r\times n_{\theta}$ points (cell centers) labelled $(i \ j)$, with $i=1\dots n_r$ and $j=1\dots n_{\theta}$. Cell interfaces between neighboring cells are denoted as $(i+\half \ j)$ for the radial ones and  $(i \ j+\half )$ for the angular ones. Hence, indices $(i+\half \ j+\half )$ denote cell edges in the $\varphi$ direction. In order to implement the CT scheme we do not define the poloidal magnetic field cell-centered but at the interfaces, i.e $B^{* r}_{i+\half \ j}$ and $B^{* \theta}_{i j+\half}$. However, as we assume axisymmetry, we use a cell-centered toroidal field $B^{*\varphi}_{ij}$  because this component does not play any role in the CT scheme. Therefore, we consider hereafter the poloidal components of the magnetic field only. With such considerations the above equations for the magnetic flux time evolution lead to equations for the magnetic field evolution (CT scheme). The poloidal components read:
\begin{eqnarray}
\left.\frac{\partial B^{*\, r}} {\partial t}\ \right|_{i+\half\ j}&=&
  \frac{\left[E^*_{\varphi} \ \Delta \hat{l}_{\varphi}\right]_{i+\half \ j-\half}
  - \left[E^*_{\varphi}\ \Delta \hat{l}_{\varphi}\right]_{i+\half \ j+\half} }
  {\Delta \su_{r \ i+\half \ j} }, \\
\left.\frac{\partial B^{*\, \theta}}{\partial t}\ \right|_{i \ j+\half} &=&
  \frac{\left[ E^*_{\varphi} \ \Delta \hat{l}_{\varphi}\right]_{i+\half \ j+\half}
  - \left[ E^*_{\varphi}\ \Delta \hat{l}_{\varphi}\right]_{ i-\half \ j+\half} }
  {\Delta \su_{\theta \ i \ j+\half} }.
\end{eqnarray}
The total magnetic flux through the cell interfaces is given by
\begin{eqnarray}
\Phi_{{\rm T} \ i\ j} &=&
\Phi_{r \ i+\half \ j} - \Phi_{ r \ i-\half \ j} 
+ \Phi_{\theta \ i\ j+\half}- \Phi_{\theta \ i \ j-\half} ,
\end{eqnarray}
where we have taken into account that the total flux in the $\varphi$ direction is zero owing to the axisymmetry condition. The time evolution of the total magnetic flux evolved with the CT scheme satisfies by construction that $[\partial \Phi_{\rm T}/\partial t ]_{i\ j} =0$, and therefore every numerical scheme constructed on the basis of the CT scheme will conserve magnetic flux up to machine accuracy. If we are able to generate initial conditions which satisfy the divergence constraint, i.e. with $\Phi_{\rm T} = 0$ at each numerical cell, then, such constraint will be preserved during the numerical evolution. The way to build such initial data is explained below.

The next step is to choose a discretization of the integrals by making explicit the values of the interface surfaces and of the edge lengths
\begin{eqnarray}
 \hspace{-1.1cm}
 &\Delta \hat{l}_{r \ i \ j + \half} = \Delta r_i,\quad
 &\Delta \su_{r \ i+\half \ j} 
   = - r^2_{i+\half} \ \Delta (\cos{\theta})_j \ \Delta \varphi,
 \\
 \hspace{-1.1cm}
 &\Delta \hat{l}_{\theta \ i+\half \ j} = r_{i+\half} \Delta\theta_j, \quad
 &\Delta \su_{\theta \ i \ j+\half}
   = \frac{1}{2}\sin{\theta_{j+\half}} \ \Delta r^2_i \ \Delta\varphi,
 \\
 \hspace{-1.1cm}
 &\Delta \hat{l}_{\varphi \ i+\half \ j + \half} =
   r_{i+\half} \sin{\theta_{j+\half}} \Delta \varphi, &\quad
 \Delta \su_{\varphi \ i \ j} 
   =\frac{1}{2} \ \Delta r^2_i \ \Delta \theta_j,
\end{eqnarray}
where 
$\Delta r_i \equiv  r_{i+\half} - r_{i-\half} $, 
$\Delta \theta_j \equiv \theta_{j+\half} - \theta_{j-\half}$, 
$\Delta r^2_i \equiv r^2_{i+\half} - r^2_{i-\half}$, 
$\Delta (\cos{\theta})_j \equiv \cos{\theta_{j+\half}} -\cos{\theta_{j-\half}} $, 
and $\Delta \varphi$ is arbitrary due to axisymmetry (hence, the cell size on the $\varphi$ direction does not play any role in the numerical scheme). Taking into account the above expressions  the evolution equations for the poloidal magnetic field read
\begin{eqnarray}
 \hspace{-0.2cm}	
\left.\frac{\partial B^{*\, r}}{\partial t}\ \right| _{i+\half\ j} &=&\frac{
  \sin{\theta_{j+\half}} \ E^*_{\varphi \ i+\half \ j+\half}
  -\sin{\theta_{j-\half}} \ E^*_{\varphi \ i+\half \ j-\half}}
  {r_{i+\half \ j} \ \Delta (\cos{\theta})_j} ,
  \label{eq:ct_r}
\\
 \hspace{-0.2cm}
\left.\frac{\partial B^{*\, \theta}}{\partial t} \ \right|_{i \ j+\half} &=&
  2 \ \ \frac{r_{i+\half} \ E^*_{\varphi \ i+\half \ j+\half} 
  - r_{i-\half} \ E^*_{\varphi \ i-\half \ j+\half} }
  {\Delta r^2_i} ,
  \label{eq:ct_theta}
\end{eqnarray}
which are used in the numerical code to update the magnetic field. The only remaining aspect is to give an explicit expression for the value of $E^{*}_{\varphi}$. A practical way to calculate $E^*_{\varphi}$ from the numerical fluxes in the adjacent interfaces~\cite{balsara99} is
\begin{eqnarray}
 \hspace{-2.1cm}	
E^*_{\varphi \ i+\half \ j+\half} = - \frac{1}{4} & &\left[
(\mb{F}^r)^{\theta}_{i \ j+\half}
+ (\mb{F}^r)^{\theta}_{i+1 \ j+\half}
- (\mb{F}^\theta)^{r}_{i+\half \ j}
- (\mb{F}^\theta)^{r}_{i+\half \ j+1}
\right ],
\end{eqnarray}
where the fluxes are obtained in the usual way by solving Riemann problems at the interfaces. The combination of the CT scheme and this way of computing the electric field is called the flux-CT scheme. It is used in all numerical simulations reported in this paper.

%%%%%%%%%%%%%%%
\subsection{The metric solver}
%%%%%%%%%%%%%%%

The CFC metric equations are 5 nonlinear elliptic coupled Poisson-like equations which can be written in compact form as  $\hat{\Delta} \mb{u} (\mb{x}) = \mb{f} (\mb{x}; \mb{u} (\mb{x}))$, where $ \mb{u} = u^k = (\phi, \alpha \phi, \beta^j) $, and $ \mb{f} = f^k $ is the vector of the respective sources. These five scalar equations for each component of $ \mb{u} $ couple to each other via the source terms that in general depend on the various components of $ \mb{u} $. We use a fix-point iteration scheme in combination with a linear Poisson solver to solve these equations. Further details on such metric solver can be found in \cite{cerda05,dimmelmeier_05_a}.

%%%%%%%%%%%%%%%%%%%%%%
\section{Initial magnetic field configurations}
%%%%%%%%%%%%%%%%%%%%%%

Since the CT scheme only preserves the value of $\vec{\hat{\nabla}}\cdot\vec{B}^*$ but does not impose the divergence constraint of the magnetic field itself, we have build initial conditions which satisfy such condition. To do this we calculate the initial magnetic field from a vector potential $\vec{A}^*$, such that $\vec{B}^* = \vec{\hat{\nabla}} \times \vec{A}^*$. We choose the following numerical discretization of this equation 
\begin{eqnarray}
 \hspace{-0.4cm}	
  B^{*r}_{i+\half \ j} &= 
  -\frac{1}{r_{i+\half}}
  \frac{\sin{\theta_{j+\half}} \ A^*_{\varphi \ i+\half \ j+\half  } 
    - \sin{\theta_{j-\half}} \ A^*_{\varphi \ i+\half \ j-\half} }
       {\Delta{(\cos{\theta})}_j}, \\
 \hspace{-0.4cm}
  B^{*\theta}_{i\ j+\half} &= -2
  \frac{r_{i+\half} \ A^*_{\varphi \ i+\half \ j+\half  } 
    - r_{i-\half} \ A^*_{\varphi \ i-\half \ j+\half} }
       {\Delta r^2_i}.        
\end{eqnarray}
which warrants that the numerical magnetic flux $\Phi_{\rm T}$ over a cell is zero (up to round-off error) in the specific numerical representation of our CT scheme. 

For our code tests and core collapse simulations we use two possible poloidal magnetic field configurations as initial conditions: a) The homogeneous ``starred'' magnetic field, in which $\vec{B}^*$ is constant and parallel to the symmetry axis, and b) the magnetic field generated by a circular current loop of radius $r_{\rm mag}$ \cite{jackson62}. Note that in both cases, we employ the ``starred'' magnetic field, since the divergence constraint is valid for this quantity when computed with respect to the flat divergence operator. In this way we can extend any analytic prescription for the magnetic field given in  flat spacetime in an easy way. Note that in the presence of strong gravitational fields the magnetic field $\vec{B}$ is deformed with respect to $\vec{B}^*$ due to the curvature of the spacetime, although the divergence constraint is automatically fulfilled.

%%%%%%%%%%
\section{Code tests}
%%%%%%%%%%

We have designed several tests in order to check the accuracy of our numerical code when solving the induction equation with the numerical methods described in the previous sections. The ``toroidal test'' is designed to assess the ability of the code to maintain equilibrium magnetic field configurations (labelled TTA and TTB) and to correctly compute the amplification of the toroidal magnetic field as it is wound up by a rotating fluid (TTC). On the other hand, the ``poloidal test" (PT) is designed to check whether the code can correctly compute the compression of the poloidal magnetic field in a spherical collapse, and its ability to handle the presence of radial shocks.

%%%%%%%%%%%%%
\subsection{Toroidal test }
%%%%%%%%%%%%%

%
\begin{SCfigure}[][t!]
\resizebox{0.5\hsize}{!}{\includegraphics*{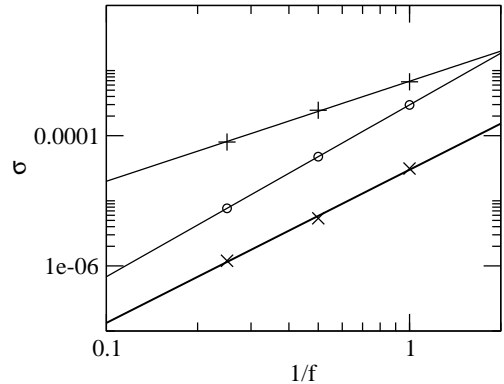}}
  \caption{Global error in the toroidal magnetic field, $\sigma$, after a time evolution of 1 ms for tests TTA ($\times$), TTB ($\opencircle$),  and TTC ($+$) as a function of $1/f$ for a sequence of models with grid resolutions $80\times 10$ ($f=1$), $160 \times 20$ ($f=2$) and $320 \times 40$ ($f=4$). The solid lines are the best fits of each test to a power law.
    \vspace{0.2cm}}
  \label{fig:TT}
\end{SCfigure}

We consider a rotating stationary configuration with no meridional flows, $v^{*r}=v^{*\theta}=0$, and $v^{*\varphi}=\Omega^*(r,\theta) \, r \sin{\theta}$, where $\Omega^*(r,\theta)$ stands for the rotation law. Under these conditions and in the passive field approximation, the induction equation can be integrated analytically. The solution shows that the poloidal component of the magnetic field remains constant and the toroidal component grows linearly with time as $B^{* \varphi} (t) = B^{* \varphi} (t=0) + t \,\, r\sin{\theta}  \,\vec{B}^* \cdot  \vec{\hat{\nabla}} \Omega^*$. In order to test whether the numerical code is able to recover this solution, we investigate three particular cases. For all three tests we consider a non-evolving fluid of constant density $\rho=10^{14}$ g cm$^{-3}$ filling the computational domain, and an equally-spaced grid in both the angular and radial directions. The outer radial boundary is located at $20$ km. We also assume a static background spacetime with a flat metric. We note that our choice of units is arbitrary since the background metric is flat. However, we maintain the adopted units to keep a link with the scenario for which the code is prepared, namely core collapse.Test TTA  consists in a rigidly rotating fluid, $\Omega^* =\Omega^*_c$, and the magnetic field generated by a circular current loop of radius $r_{\rm mag} = 6$ \,km. Test TTB includes a homogeneous magnetic field and a differentially rotating fluid with a rotation law given by $\Omega^* (r, \theta)= (A^2 \Omega^*_c)/(A^2+(r \sin{\theta})^2)$, with $A=6$ km. In both cases the analytic solution corresponds to a non-evolving toroidal magnetic field because $\vec{B}^* \cdot  \vec{\hat{\nabla}} \Omega^*=0$. The third test, TTC, consists in a homogeneous magnetic field and a fluid satisfying a rotation law depending on the radius $\Omega^* (r)= (A^2 \Omega^*_c)/(A^2+r^2)$, and $A=6$ km. In this case the analytical solution is given by 
\begin{equation}
B^{*\varphi}_{\rm (th)} (t) = - B^*_0 \Omega^*_c \frac{A^2 r^2 \, t}{(A^2+ r^2)^2} \, \sin{2 \theta}.
\label{eq:TTC_eq}
\end{equation}
In all three cases the central angular velocity $\Omega^*_c$ is such that the center rotates 10 times during the complete evolution, namely  1 ms. Once this time is reached we compute the local error $\sigma_{ij}\equiv|B^{\varphi} - B^{\varphi}_{\rm (th)}|$ between the analytical solution $B^{\varphi}_{\rm (th)}$ and the numerical one, and the global error $\sigma$ calculated as the L2-norm of the local error.

\begin{figure}[t!]
\flushright
  \resizebox{0.85\hsize}{!}{\includegraphics*{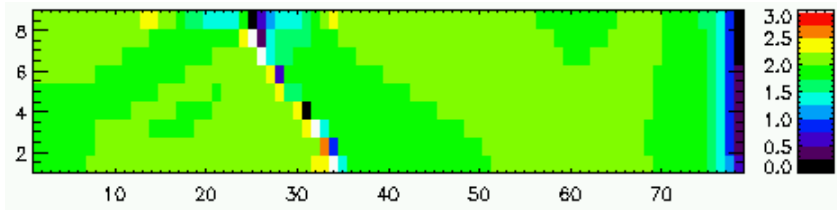}}
  \caption{Local order of convergence (color coded) for the TTA  test after a total time evolution of $1$ ms. White color is used for values larger or equal than $3.0$.  The axes represent the number of cells of the reference grid in the radial  and angular direction.} 
  \label{fig:TTA_nij}
\end{figure}

If we consider a reference (low resolution) grid $n_{r\,{\rm ref}} \times n_{\theta\,  {\rm ref}}$ with an error $\sigma_{\rm ref}$ and a subsequent grid with a factor $f$ higher resolution, i.e. $n_{r} = f n_{r\,{\rm ref}}$ and $n_{\theta} = f n_{\theta\,{\rm ref}}$, then the error on the new grid, $\sigma_f$, is related to the error on the reference grid as $\sigma_{f} = \sigma_{\rm ref} (1/f)^N$, where $N$ is the order of the numerical scheme. By fitting the values of $\sigma_f$ as a function of $1/f$ we can easily calculate the convergence order $N$. For our test simulations we consider a reference grid of $80\times 10$ cells ($f=1$) , and higher resolution grids of $160\times 20$ ($f=2$) and $320\times 40$ ($f=4$) cells, respectively. Figure~\ref{fig:TT} shows the resulting $\sigma_f$ for the three tests versus $1/f$. Our results show that (i) the order for the TTC test ($N=1.54$, obtained by fitting the data to a power law) is smaller than for the TTA and TTB tests ($N=2.35$ and 2.64, respectively), and (ii) the order for cases TTA and TTB is $N>2$, and hence higher than the theoretical expectation (second order given by the time and space discretization order). 

The main difference between the case TTC and the cases TTA and TTB, which explain these results, is that in the first case there is a component of the magnetic field, $B^{* \varphi}$, which grows linearly in time, while in the other two cases no components evolve. Hence, the order of convergence for tests TTA and TTB is higher than for test TTC. We suspect that this is due to more precise numerical cancellations in Eqs.~(\ref{eq:ct_r}) and (\ref{eq:ct_theta}) (which have $E^*_{\varphi}\ne0$) in tests TTA and TTB. This can be explained by computing the {\it local} order of convergence, i.e. the order obtained when computing the errors of each numerical cell, $\sigma_{ij}$, instead of the global error $\sigma$. The results for test TTA are displayed in figure~\ref{fig:TTA_nij} (similar plots can be obtained for the other two cases). It can be seen that at some particular grid zones the order of convergence is larger than two, while at most locations  it remains around two. 

\begin{figure}[t!]
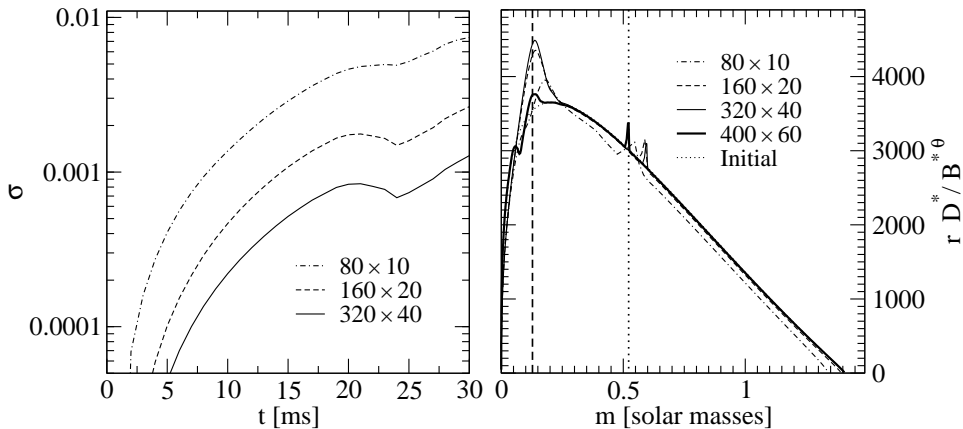

  \flushright
   \resizebox{!}{0.43\hsize}{\includegraphics*{figure4a.eps}}
 \resizebox{!}{0.43\hsize}{\includegraphics*{figure4b.eps}}
  \caption{Results for the poloidal test (PT) for different grid resolutions. The left panel shows the global error as a function of time during the infall phase, while the right panel shows the quantity $r\, D^*/B^{\theta *}$ versus $m$ at $t=35$ ms. The location of the inner core of the PNS (dashed vertical line) and the location of the shock  (dotted vertical line) for the highest resolution model are also shown.}
  \label{fig:PTA}
\end{figure}

We note that the computation of the initial magnetic field for test TTA is not analytic but the result of an expansion with infinite terms. This expansion is computed to a required level of accuracy, typically smaller than the accuracy required during the evolution in the routine to recover the primitive variables from the state vector. For a particular direction, however, close to the center of the ``peculiar" cells in figure~\ref{fig:TTA_nij}, the solution needs infinite terms, and the resulting solution is thus less accurate at these points. The physical meaning of this direction in the equatorial plane is that it coincides with the radius of the current loop which generates the initial magnetic field.

%%%%%%%%%%%%%
\subsection{Poloidal test} 
%%%%%%%%%%%%%

Next, we consider a test in which only radial velocities of the fluid are allowed, i.e. $v^{* r}\ne 0$ and $v^{* \theta} = v^{* \varphi}=0$. We also consider initially a purely poloidal magnetic field. In this case, it can be easily shown from the induction equation~(\ref{eq:induction}) and the continuity equation~(\ref{eq:hydro_conservation_equation}) that the following equivalence holds in the equatorial plane
\begin{equation}
\frac{\partial}{\partial t} \left(
r \frac{D^*}{B^{* \theta}} \right) =
- v^{* r} \ \frac{\partial}{\partial r}
\left( r \frac{D^*}{B^{* \theta}}\right),
\end{equation}
which is an advection equation for $r \, D^*/B^{* \theta }$ with velocity $v^{* r}$ at the equator. Since only radial velocities are allowed, it is possible to define a Lagrangian coordinate system in which the value 
$r\,D^*/B^{\theta *}$ does not change with time.  The easiest way of checking this condition is to calculate 
$\left[r\,D^*/B^{\theta *}\right] (m)$, i.e. as a function of the mass enclosed within a radius $r$ defined as $m (r) \equiv 4 \pi \int_0^r r'^2 dr' D^*(r')$. This quantity should remain constant in time, provided the magnetic field is correctly evolved with our numerical code.

\begin{SCfigure*}[4][t]
\includegraphics*[width=.5\textwidth]{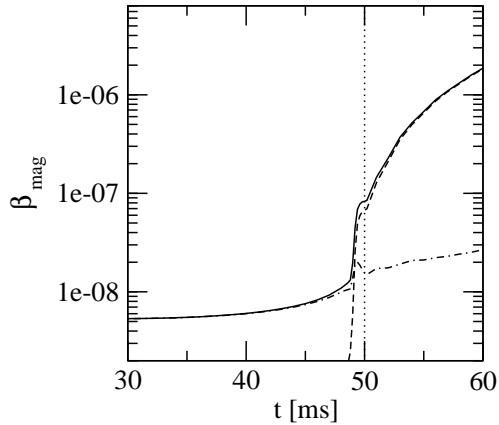}
  \caption{Time evolution of the  magnetic energy parameters  $\beta_{\rm mag}$ (solid line), $\beta_{\varphi}$ (dashed line), and $\beta_{\rm polo}$  (dot-dashed line) during magneto-rotational collapse. 
 The vertical dotted line indicates the time of bounce. \vspace{0 cm}} 
  \label{fig:beta_mcollapse}
\end{SCfigure*}

To test this fact we perform simulations of spherical collapse in which the above conditions are satisfied.
The initial model is a spherical equilibrium 4/3-polytrope (TOV) with central density $\rho_c=10^{10}$ g cm$^{-3}$ and a homogeneous (starred) magnetic field. We use a hybrid EOS (see \cite{dimmelmeier_02_a}). We induce the collapse by reducing the initial adiabatic index to $\gamma_1=1.28$. As the EOS stiffens at nuclear matter density the star bounces ($t_{\rm b} = 30$ ms) and a shock forms, which travels outward. We calculate the global error $\sigma$ in the collapse phase as the L2-norm applied to the differences in the quantity $\left[r\,D^*/B^{\theta *}\right] (m)$ between its initial value and its value at subsequent time steps. In the left panel of figure~\ref{fig:PTA} we show the evolution of such error during the collapse for different $(r,\theta)$-grid resolutions,  equally-spaced in the angular direction and logarithmically spaced in the radial direction,  except for the inner $20$ radial grid points which are equally-spaced. In each case the errors are below $1\%$, even for the coarsest grid, and the computed order of convergence during the collapse (at $t=20$~ms) is 1.41.

In order to check our numerical code after core bounce we plot in the right panel of figure~\ref{fig:PTA}  $\left[r \,D^*/B^{\theta *}\right] (m)$ at time $t=35$ ms, for the three different grid resolutions. The initial profile is shown by the dotted curve. The effect of the travelling shock is seen as a small spike, that becomes narrower as the resolution is increased. On the other hand, deviations from the initial profile occur near the inner core boundary of the PNS (where the density exceeds nuclear matter density). 
This numerical error becomes smaller as the resolution covering the inner region is increased by using a grid of $400\times60$ zones, with $100$ equally-spaced radial zones in the inner part instead of $20$. This resolution is the same as the one used in magneto-rotational core collapse simulations. It guarantees sufficient resolution in the region where the PNS forms. We have performed comparisons of the HLL approximate Riemann solver and the KT symmetric scheme, finding almost identical results (in agreement with \cite{lucas04,shibata_font05,anton06}). 

%%%%%%%%%%%%%%%%%%%%%
\section{Magneto-rotational core collapse}
%%%%%%%%%%%%%%%%%%%%%

We end the paper by discussing a representative magneto-rotational core collapse simulation. A parameter-space survey of a large sample of models will be presented elsewhere. The initial model, which we label A1B3G3-D3M0, is a general relativistic and rotating $4/3$-polytrope in equilibrium, and is the magnetized version of the purely hydrodynamical model A1B3G3 of \cite{dimmelmeier_02_a}, which is differentially and rapidly rotating. The configuration of the initial magnetic field is a circular current loop of radius $r_{\rm mag} = 400$ km with a strength at the center of the loop of  $B_{\rm c} = 10^{10}$ G. For such value the passive field approximation is adequate, since, locally, $P_{\rm mag}/P < 10^{-6}$ throughout the simulation. The collapse begins by lowering the adiabatic index to $\gamma_1=1.31$. The same hybrid EOS as in the PT test is used. We employ the numerical grid described in the previous section.

\begin{figure*}[t]
  \flushright
  \resizebox{0.5\textwidth}{!}{\includegraphics*{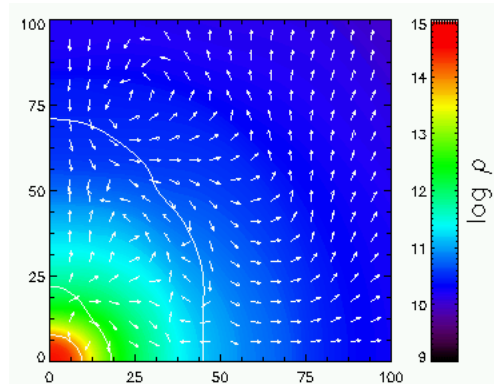}}
  \caption{Configuration of the innermost region of the star at the end of the magneto-rotational collapse of its core ($t=60$ ms). The panel shows the logarithm of density, $\log{\rho}$, in g cm$^{-3}$, the distribution of the velocity field, $v^i$ (arrows), and contours of the specific internal energy. All axes are in km.}
   \label{fig:A1B3G3_D3M0_color_bounce1}
\end{figure*}

Figure~\ref{fig:beta_mcollapse} shows the evolution of the energy parameter for the magnetic field, $\beta_{\rm mag}$, defined as the ratio of magnetic energy, $E_{\rm mag}= \frac{1}{2} \int \mathrm{d}^3 \mb{x} \, W b^2$, to the potential energy $E_{\rm pot}$ (see e.g.~\cite{dimmelmeier_02_a}). In order to analyze the growth of the magnetic field, we separate the effect of the different components of the magnetic field into $\beta_{\varphi}$, for the toroidal component, and $\beta_{\rm polo}=\beta_{\rm mag} - \beta_{\varphi}$, for the poloidal component, which are also plotted in the figure. As the collapse proceeds, and leaving aside the effects of the magneto-rotational instability (MRI), the magnetic field grows by at least two reasons: First, the  radial flow compresses the magnetic field lines, amplifying the existing poloidal and toroidal magnetic field components. Second, during the collapse of a rotating star differential rotation is produced, even for rigidly rotating initial models, as shown in numerical simulations (e.g.~\cite{dimmelmeier_02_a,ss04}). This fact can be easily explained considering that energy and angular momentum are roughly conserved in cylindrical regions (see e.g.~\cite{meier76}). Hence, if a seed poloidal field exists, the $\Omega$-dynamo mechanism acts winding up poloidal field lines into the toroidal component. This (linear) amplification process generates a toroidal magnetic field component, even from purely poloidal initial configurations. The toroidal component of the magnetic field is affected by the two effects while the poloidal field is only amplified by the first effect (radial compression). Thus, even if the initial magnetic field configuration is purely poloidal, the toroidal component dominates after some dynamical time. This can be seen  in figure~\ref{fig:beta_mcollapse} as $\beta_{\phi}$ (dashed line) grows much faster than $\beta_{\rm polo}$, particularly after bounce ($t\sim 50$ ms) when the radial compression mechanism stops. We note that as the magnetic field considered is weak enough not to affect the dynamics, the final $\beta_{\rm mag}$ is much less than unity. 

\begin{figure*}[t]
  \flushright
  \resizebox{1.0\textwidth}{!}{\includegraphics*{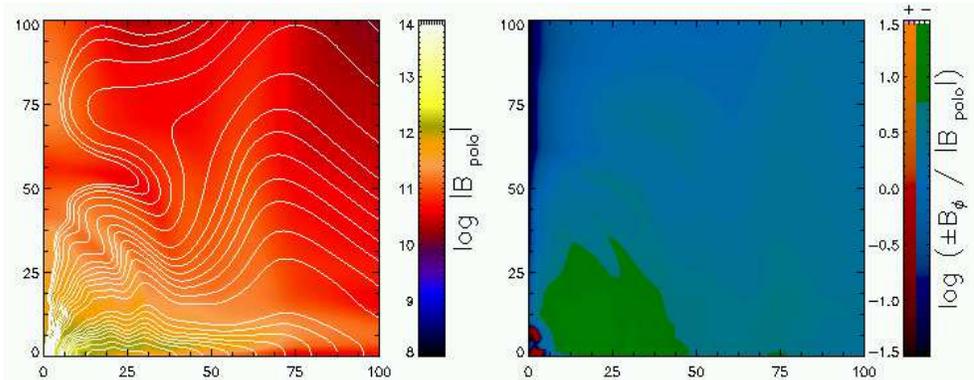}}
  \caption{Configuration of the innermost region of the star at the end of the magneto-rotational collapse of its core ($t=60$ ms). The left panel shows the logarithm of the poloidal component of the magnetic field, $\log{|B_{\rm polo}|}$, in G  and the magnetic field lines in the $r$-$\theta$ plane (lines). The right panel shows, $B^{\varphi}/|B_{\rm polo}|$. All axes are in km.}
   \label{fig:A1B3G3_D3M0_color_bounce2}
\end{figure*}

Figures~\ref{fig:A1B3G3_D3M0_color_bounce1} and  \ref{fig:A1B3G3_D3M0_color_bounce2} show the innermost 100 km of the star after core bounce.  A PNS has formed at the center with two distinct regions. The inner core of $\sim 10$ km, where nuclear matter density is reached, has a mixed poloidal-toroidal magnetic configuration, while there exists a surrounding shell extending to $\sim 50$ km, with sub-nuclear densities, that has a dominant toroidal magnetic field which grows linearly due to the $\Omega$-dynamo mechanism resulting from the strong differential rotation of this shell. From the linear growth of this component we estimate that the magnetic field is likely to reach saturation values of $\sim 10^{15}$ G on a timescale of several seconds. However, by estimating the timescale of the fastest growing unstable mode of the MRI in our simulation, we have checked that a significant fraction of the newly-formed PNS, as well as the region behind the shock at the moment of its formation, could be affected by the MRI. Such instability
can grow on dynamical timescales of several tens of ms \cite{martin}, much faster than the $\Omega$-dynamo mechanism. However, it is still an open issue~\cite{Hawley} whether, for the weak magnetic fields encountered in core collapse progenitors, the amplification of the magnetic field is going to become important for the dynamics. If that were the case, the passive field approximation would not be valid in such regions, becoming necessary a full magnetic field treatment.

%%%%%%%%%%
\section{Summary}
%%%%%%%%%%

We have presented a new general relativistic hydrodynamics code specifically designed to study magneto-rotational, relativistic, stellar core collapse. The code is built on an existing  hydrodynamics code which has been thoroughly applied in the recent past to study relativistic rotational core collapse~\cite{dimmelmeier_02_b}. It is based on the conformally-flat  approximation of Einstein's field equations  and conservative formulations for the magneto-hydrodynamics equations. As a first step towards magneto-rotational core collapse simulations, the code assumes a {\it passive} (test) magnetic field, a justifiable assumption since weakly magnetized fluids are present in this scenario . The paper has focused on the description of the technical details of the numerical implementation, with emphasis on the magnetic field module. A number of tests have been presented and discussed, and the convergence properties of the code have been analyzed. In addition, a representative magneto-rotational core collapse simulation has been presented. The amplification of the magnetic field due to the $\Omega-$dynamo mechanism has been estimated (acting on timescales of several seconds). More detailed studies of faster amplification mechanisms (namely, MRI) are necessary for initially weak magnetic fields, since they could dominate the post-bounce dynamics within a few ms and might have a major effect on the gravitational wave signal.
  
\ack{
We are grateful to Luis Ant\'on for his continuous advice during the development of the numerical code and for many stimulating discussions. Research supported by the Spanish {\it Ministerio de Educaci\'on y Ciencia} (grant AYA2004-08067-C03-01).}

\end{document}